\documentstyle[11pt,newpasp,twoside,epsf]{article}
\def\edcomment#1{\iffalse\marginpar{\raggedright\sl#1\/}\else\relax\fi} 
\reversemarginpar 

\begin{document} 
\title{Future prospects: Merging data with models}

\author{Gerard Gilmore} 
\affil{Institute of Astronomy, Madingley Road, Cambridge, UK}

\begin{abstract} 
Globular clusters are stellar dynamical systems which evolve on
stellar evolutionary and both internal and external dynamical
timescales. Quantitative comparison of cluster properties with
realistic evolutionary dynamical models is becoming feasible, and will
underpin the subject in the next few years. A few examples of the
types of analysis becoming possible are presented.

\end{abstract} 

\section{Introduction} 

Considerable progress has been reviewed at this meeting in quantifying
the luminosity function of globular cluster systems at birth, and its
evolution due to a mix of stellar evolution and environmental
effects. Even greater progress has been made in quantifying the
stellar mass function in clusters, and its spatial evolution (mass
segregation).  Cluster ages and morphologies are now available in
large numbers and many environs. The next stage is to develop and
apply full time-dependent dynamical models, including the (dark)
stellar remnants and external tides, use these to predict stellar
colour-magnitude data, and compare these models to the mix of imaging
(HST, NGST, AO/8metre, ..) and kinematic (GAIA, SIM, massive IFU
studies...) which will become available over the next few years.  In
this talk I shall illustrate the next developments beyond King models,
noting our biggest current challenges: determining the internal
kinematics of clusters, and careful studies of very young clusters, to
quantify the range of initial conditions.

\section{Dynamical models: young wine}

The GRAPE N-body machines are in operation, and already producing
fruit (Makino 2002). Available codes include very extensive treatment
of stellar evolution, and are limited largely by a lack of observational
constraints on the boundary conditions at young ages, and
(temporarily) by star numbers. Real simulations of clusters as rich as
those in the LMC are however already possible (figure 1). These
simulations illustrate both the wealth of information potentially
available in an observational colour-magnitude diagram (CMD), and how
complex is the `main sequence turn-off' in a  young cluster. Future
CMD analyses clearly must do more than fit a single isochrone from
evolutionary models of single stars in such circmstances. Similar new
opportunities for understanding exist in older clusters, especially
from careful consideration of blue stragglers.

\begin{figure}
\plotfiddle{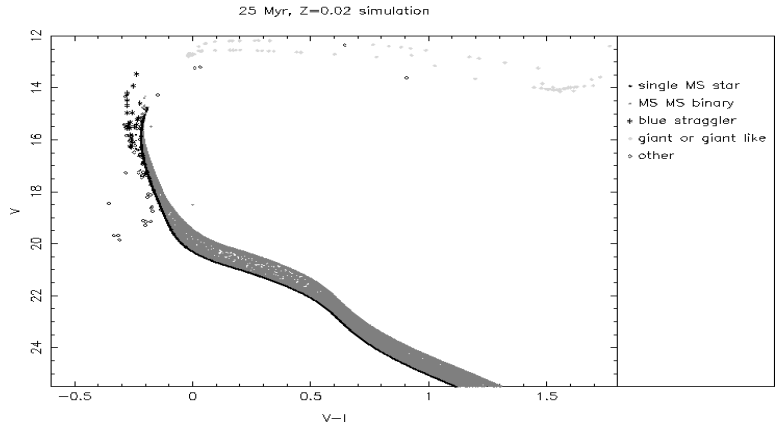}{6truecm}{270}{40}{30}{-140}{170}
\caption{Cluster CMD from a GRAPE N-body simulation, illustrating both
that stellar evolution, mass transfer, merging, etc, is now fully
included in dynamical models, and that the concept of a `main-sequence
turnoff' for a young cluster is more complex than sometimes
appreciated.  Convolution of such model clusters with a HST
observation can be compared directly with data. (from Johnson etal
2001) }
\end{figure}

\section{Real Dynamical models: cluster internal evolution}

The evolution of a (globular) star cluster is a function of stellar
evolution, as noted above, and also internal and external dynamical
effects. The most sensitive internal parameters have long been
suspected to be initial density, fraction of hard binaries, and the
stellar IMF. Recent results have highlighted the irrelevance of some
of these, and the need for more understanding of the others. 

Figure 2
presents a recent example of a relevant study, determing the core
radius-age distribution for clusters in the Galactic satellites (LMC,
SMC, Fornax, Sgr). It is apparent from figure~2 that there is a
substantially greater dispersion in core radius size in older clusters
than exists in young clusters. Does this indicate a range of initial
conditions, with (by chance?) few large young clusters forming today,
or evolution, with core radius increasing with age? Cluster size is
determined by the pressure-gravity balance, so core radius evolution
can be due either to a pressure change (increased stellar
velocity dispersion) at approximately constant total potential, or to
decreasing binding energy, from mass loss, at approximately constant
velocity dispersion, or some combination of both. 

\begin{figure}[ht!] 
\vskip 8truecm
\caption{The top left figure shows the cluster core radius vs age
relationship, derived from HST imaging data for 53 LMC clusters by
Mackey and Gilmore (2002). The 3 images show a typical young cluster
(lower left:NGC2156) and two old clusters (NGC1916, NGC1841- top), similar
in every way except for their clearly different core radii.
THIS FIGURE AVAILABLE IN gif FORMAT
}
\end{figure}

A recent series of HST studies and N-body has addressed these
issues. The easiest possible cause to check is the stellar IMF. A very
top-heavy IMF will lead to more stellar evolution-induced mass loss
than will a shallow IMF, and consequently very different contributions
to the natural evolutionary change in gravitational potential depth.
The left panel of Figure 3 illustrates the effect different IMF slopes
would have on core radius evolution. It also identifies three pairs of
clusters, of similar age and very different core radius, whose IMFs
were determined from deep HST imaging.

The IMFs were indistinguishable
-- an apparently universal observation in studies of globular cluster
IMFs --, excluding this as a contribution to core radius differences
between the clusters (de Grijs etal 2002). The right panel of figure 3
addresses the initial condition question: might the core radii be an
extreme function of stellar mass, so that comparison of young and old
at similar apparent magnitudes, but effectively at very different
stellar masses, is misleading. That is, might there be an extreme
range of initial mass segregations. This figure shows core radii of
the six clusters studied for their IMFs determined at the same low
mass (0.8M{$_{\sun}$}). While all core radii are larger (there {\sl
is} substantial early/initial mass segregation in clusters), the increased
range of core radii with increased age is still apparent.

\begin{figure}
\plottwo{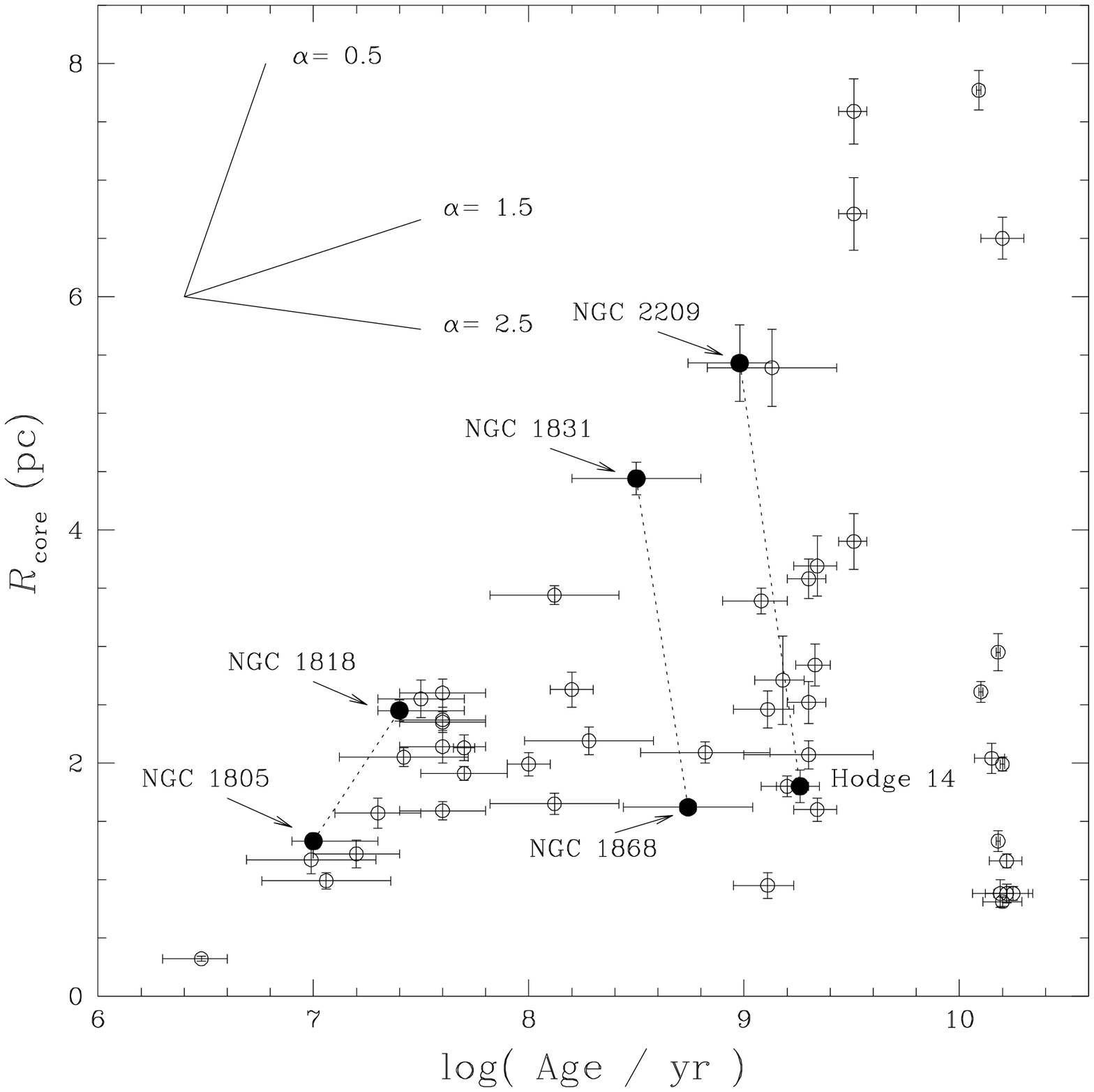}{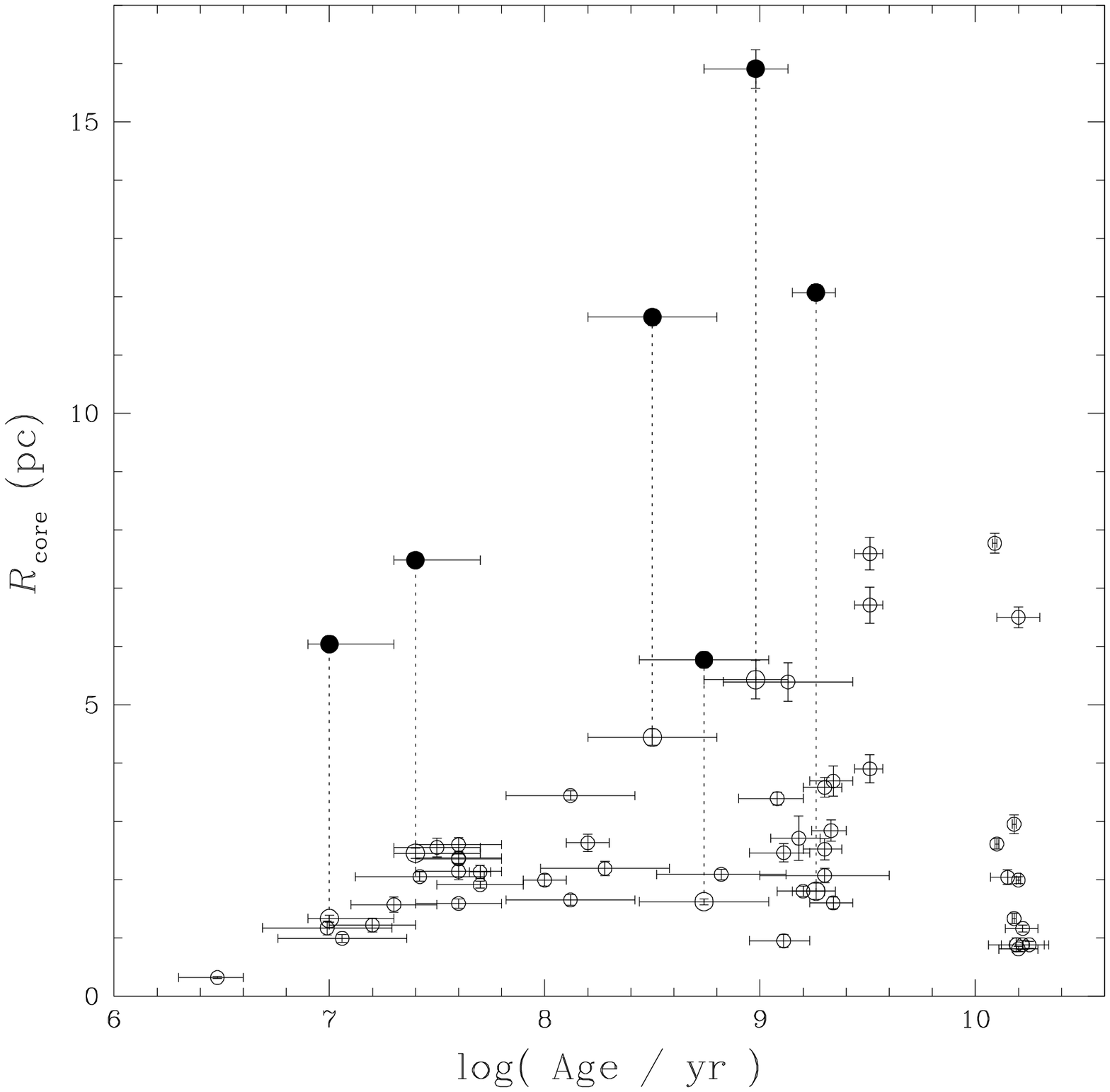}
\caption{The LMC core radius-age relation, with the direction of
evolution resulting from different stellar IMF slopes indicated (left
panel). The three pairs of clusters studied to limit IMF changes are
indicated. Right: the effect of mass on core radius for the 3 pairs of
studied clusters. core radii for 6 clusters determined for turn-off
stars and stars of 0.8Msun are connected by dots.  As expected from
mass segregation, clusters are larger when low-mass stars are used as
tracers, but the form of the age-core radius relation is
retained. (from deGrijs etal 2002)}
\end{figure}

\subsection{ Stellar IMFs in clusters}

It is worth noting here a recent study of an extremely diffuse stellar
`cluster', the UMi dSph galaxy (Wyse etal 2002). This galaxy has
similar metallicity, age and stellar mass to a large old globular
cluster, but has  a vastly lower stellar density, and a vastly higher
dark matter content. Thus, comparison of
its stellar IMF with that of similar metallicity and age clusters
provides a direct test of possible environmental effects on the IMF.
Comparison of the UMi luminosity function with that of its two `twin'
clusters, M15 and M92, is shown in Figure 4. The mass functions are
clearly very similar. This result supports an increasing number of
studies which show that the IMF at low masses behaves as a universal
constant. While neither expected nor understood, this does remove a
potentially troublesome free variable from dynamical studies.

\begin{figure}
\plotfiddle{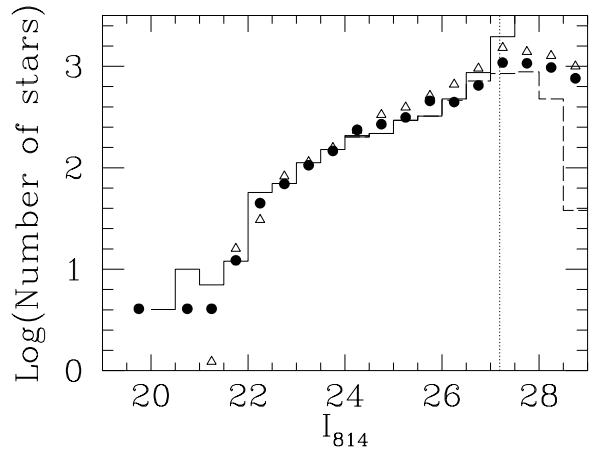}{6truecm}{270}{30}{30}{-140}{170}
\caption{ The stellar luminosity function of the UMi dSph galaxy,
compared to those for two globular clusters (M15, M92) of similar age
and metallicity. The stellar low mass IMF is manifestly similar, even
though these systems cover the full stellar density range in the
Universe. from Wyse etal (2002)}
\end{figure}

\subsection {Dynamical models: what is the answer?}

If mass loss variations are not the origin of the range of cluster
sizes, might it be a range of heating processes? Current studies
suggest not. N-body simulations of the effects of the two most
important processes, primordial hard binaries and external tidal
effects, have recently been completed.  Primordial hard binaries, a
dominant heat source in cluster dynamical evolution, are certainly
important. However, hard binaries act more as an isothermal reservoir,
preventing core collapse, than as a heat engine, systematically
increasing the internal dispersion. External tides are also surely
important in cluster evolution, and lead to evaporation, as the
several recent dicoveries of tidal tails illustrate. However, external
tides act on the outer parts of clusters more than the cores: they are
inefficient at heating core radii (figure 5) in realistic potentials
like those of the galactic satellites.

\begin{figure}
\plotfiddle{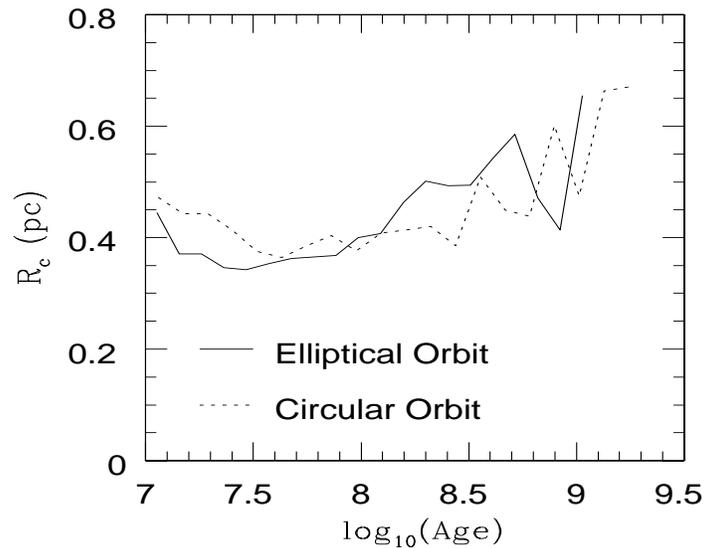}{7truecm}{0}{100}{80}{-150}{-350}
\caption{ Core radius evolution for N-body simulations of clusters
similar to those in the LMC, on orbits about a point-mass representing
the LMC. (Wilkinson etal in prep)}
\end{figure}

Unsatisfactorily, we are left with initial conditions: large clusters
formed large, possibly due to differences in star formation
efficiency, and stayed that way, until evaporated by external tides.
All is not gloom: this does lead to new possibilities for the analysis
of cluster system evolution. If the large clusters are weakly bound,
they become a tracer of the dynamical history of their environs. If
they are not, possibly exotic processes are involved. We consider each
option in turn.

\section{Exotic dynamics}

Do globular clusters host black holes? Observationally, this question
remains open. However, massive black holes are apparently ubiquitous
in the dense nuclear stellar clusters of galaxies, and low-mass black
holes are well-known in X-ray astronomy. Black holes can and do exist,
so it is of interest to investigate how they might evolve in a
cluster. The prediction is not obvious: in principle, the two holes
merge after orbital decay dominated in its late stages by
gravitational radiation. But the centre of a dense star cluster is an
active environment, with continual dynamical interactions. Do these
continual star-binary interactions encourage or delay black hole
coalescence? Specifically, do two low-mass black holes merge (and
perhaps seed an eventual super-massive nuclear black hole), or do
they act as an extreme hard binary, and disrupt the cluster? Or both?

The key parameter for binary coalescence is the binary orbital
eccentricity: GR radiatiative orbital decay is rapid only if the
orbital eccentricity approaches unity. But the orbit is continually
scattering/scattered by cluster stars. Which process dominates? Some
recent GRAPE calculations by Sverre Aarseth of merging clusters, each
of which contains a black hole, are shown in Figure 6: while this is
not always the case, in these examples a merger occurs before
effective cluster disruption. A side-effect is also apparent: the
stars scattered out of the cluster by the black hole binary have a
considerable dispersion: no cold tidal-tail would be seen in such a
case. Perhaps cold tidal tails are the strongest evidence that
globular clusters do not host more than one black hole.

\begin{figure}
\plotfiddle{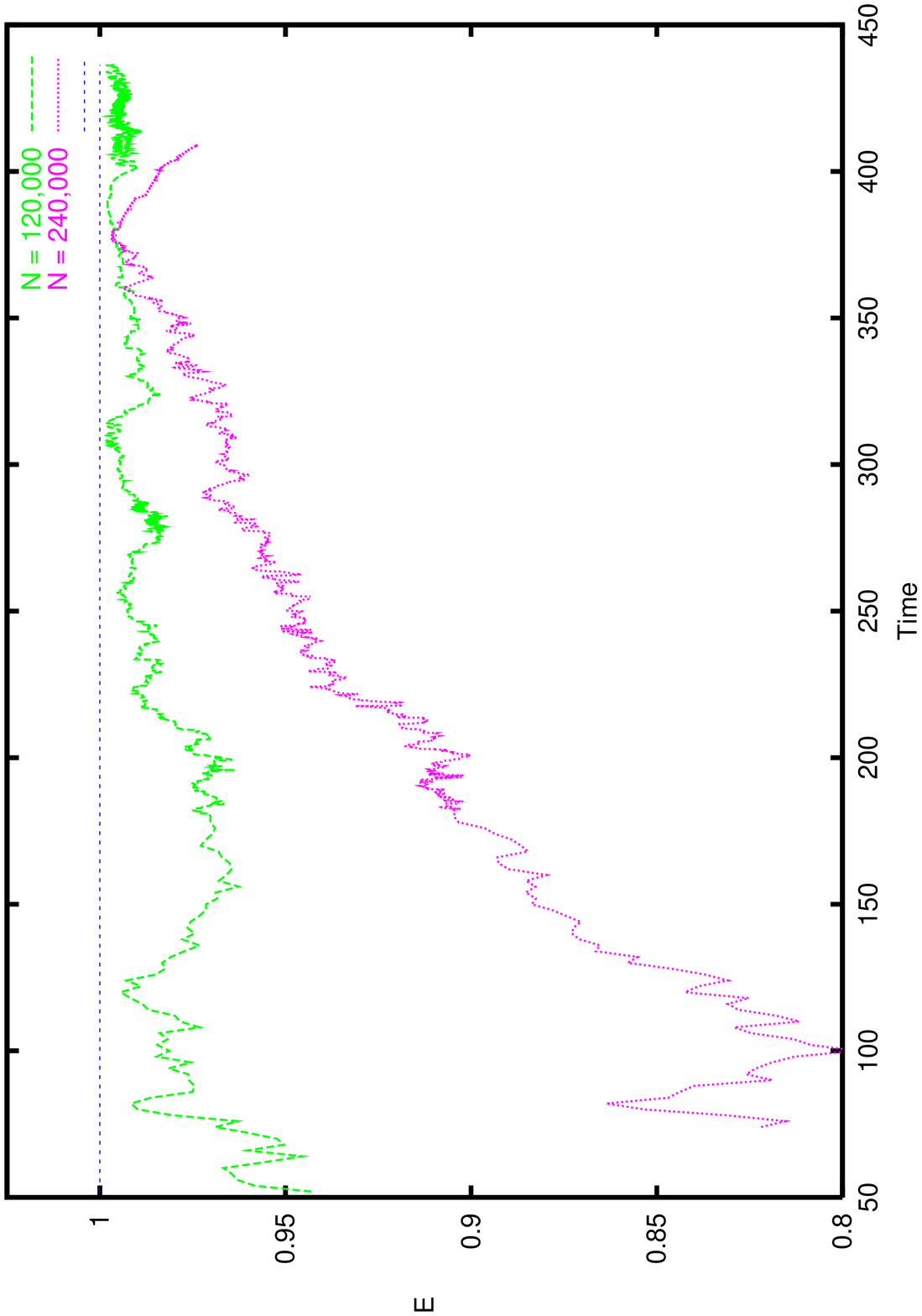}{4truecm}{270}{40}{25}{-180}{150}
\plotfiddle{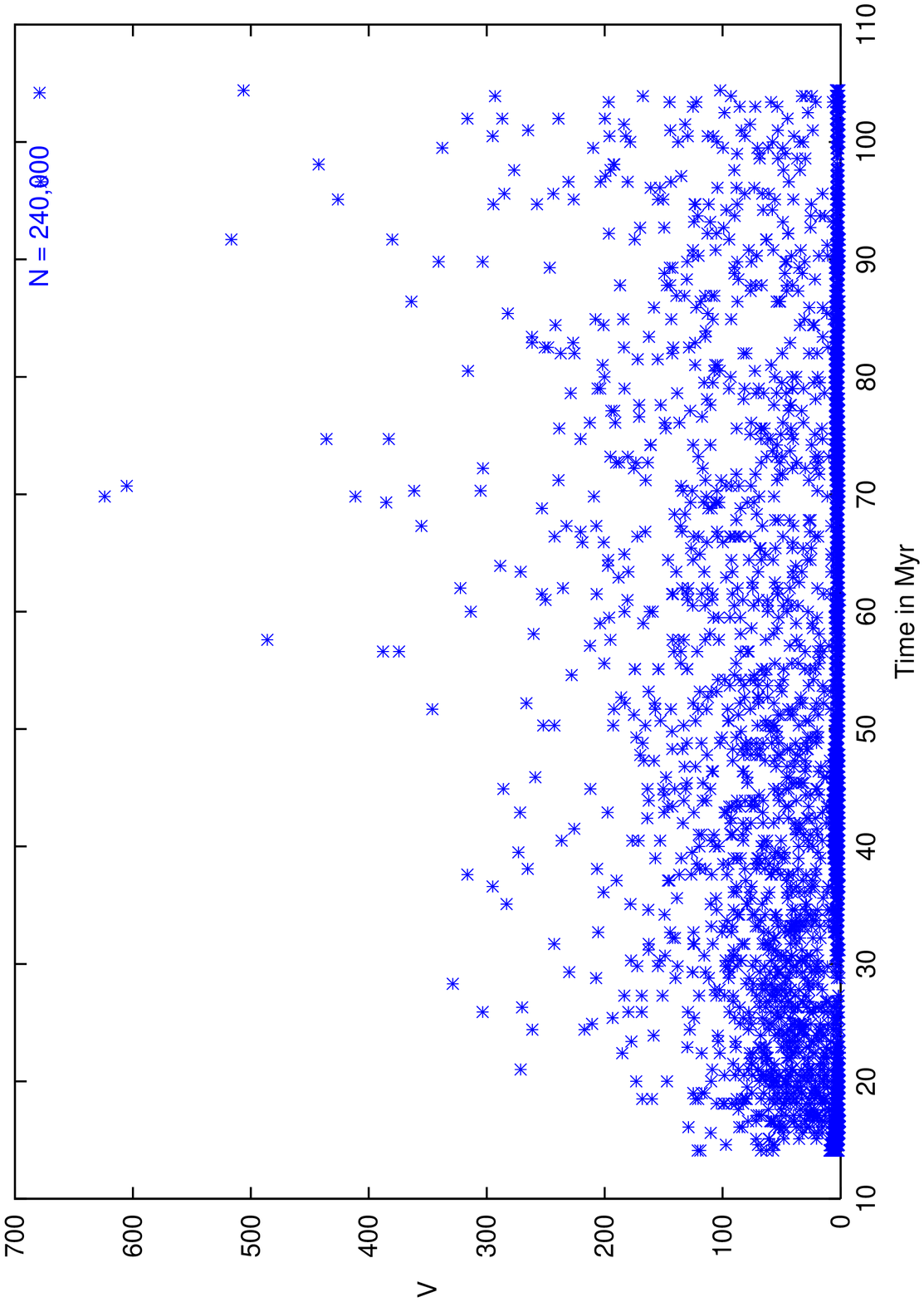}{4truecm}{270}{40}{25}{-180}{150}
\caption{N-body simulation (by Sverre Aarseth) of the evolution of a
merger of two clusters each containing an intermediate mass black
hole. Clusters of 120K and 240K stars are shown. This rigorous GR
calculation shows that the two black holes can merge into a single
central hole (eccentricity=1: top panel) before the cluster is
effectively unbound. The bottom panel shows that many stars are
ejected with well above the escape velocity: this evolutionary path
would not leave a detectable `tidal debris stream' after cluster
disruption.}
\end{figure}

\section{Dynamics and the environment}

Large weakly-bound clusters are readily damaged by tides. So if
they exist a gentle history may be inferred. Do they exist, and where?
Figure 7 illustrates the type of analysis which is becoming possible,
combining dynamical models with morphological studies.

\begin{figure}
\plottwo{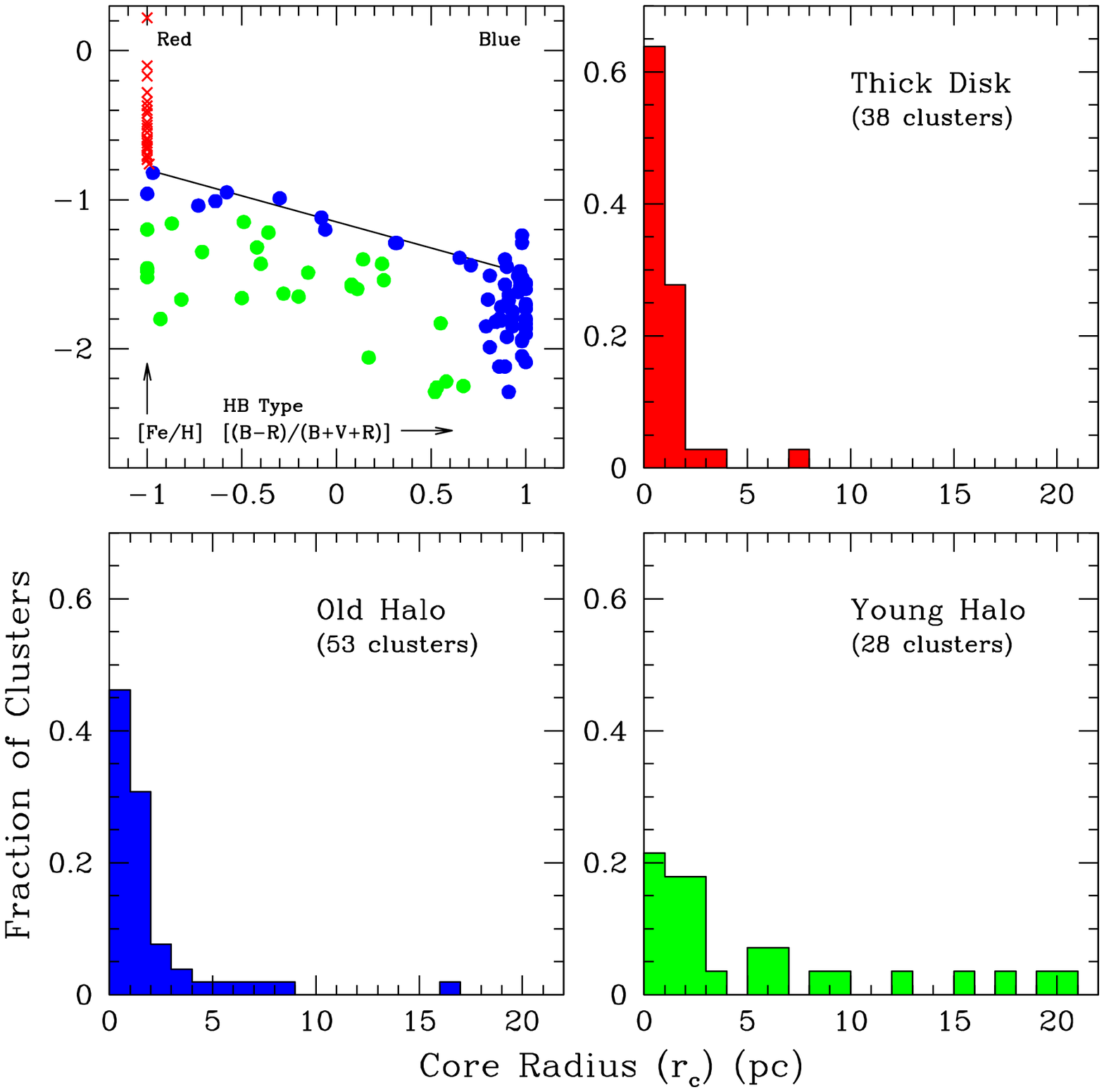}{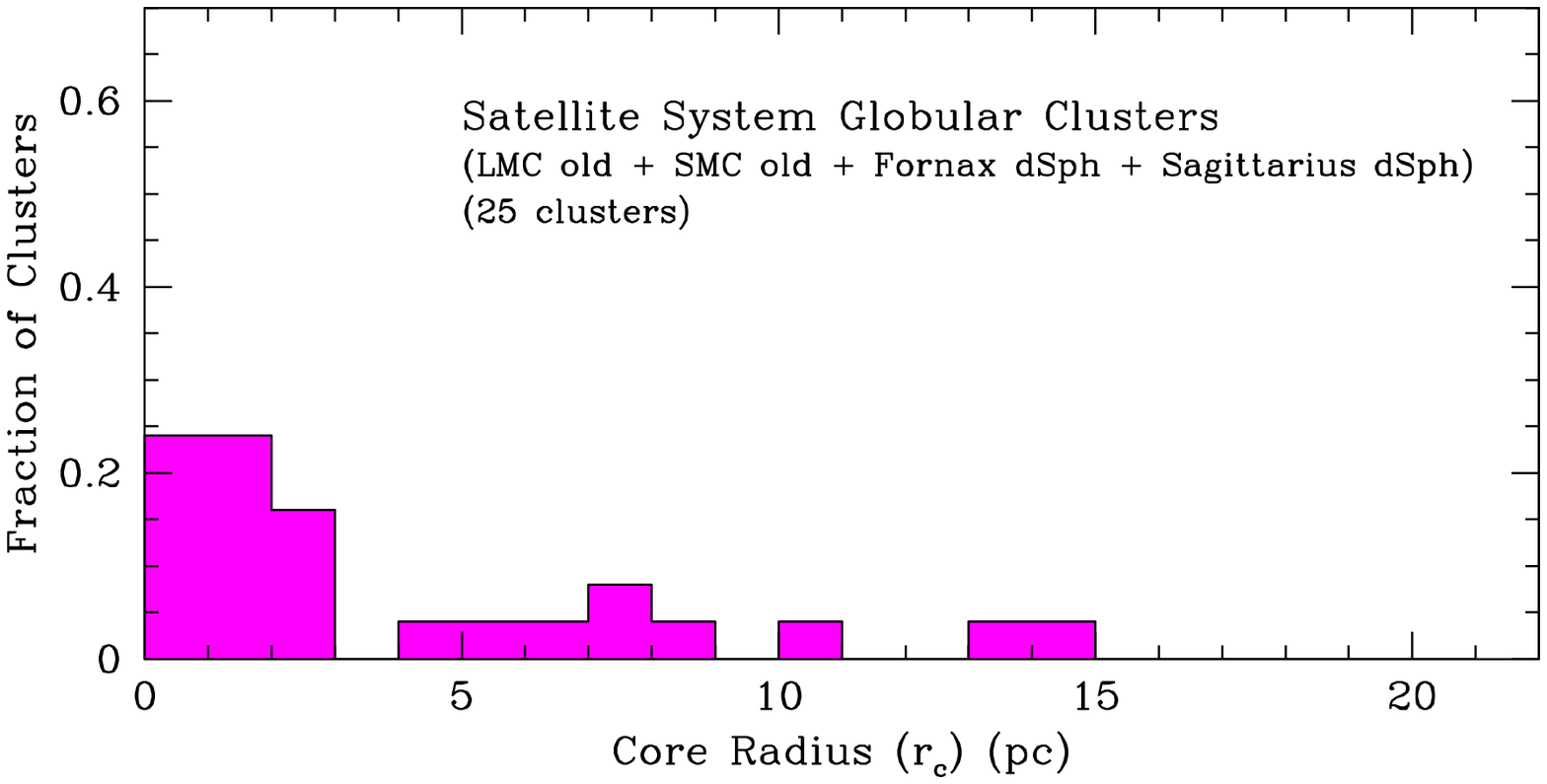}
\caption{LHS: Top left: [Fe/H] vs HB-type, defining the three subsystems
following Zinn -- bulge/thick disk (top left), `old halo', (right, and
on the solid line), and `young halo' (below the line). Other panels:
core radius distributions for the three subsystems. The young halo
distribution clearly differs from the other two, consistent with a
different dynamical history. RHS: core radii distribution for old
clusters from the 4 Galactic satellites, LMC (14), SMC (1), Fornax (5)
and Sgr (5). The distribution clearly is similar to that of Galactic
`young halo' clusters. (from Mackey \& Gilmore, this meeting)}
\end{figure}

Figure 7 illustrates the significant correlation between core radius
size and stellar horizontal-branch (HB) morphology seen in Galactic
globular clusters. This does not require a direct connection: one
might prefer to explain HB morphology (at least in part) as an age
effect. It does allow one to deduce that the `young halo' clusters
defined by HB morphology have not lived long in a harsh tidal
field. Interestingly, the right panel shows that a very similar
distribution of core radii is seen in satellite cluster systems. The
tidal field in these systems is gentle: no disk or bulge shocking. So
one may speculate that the satellites display a core size distribution
which is modified from the `primordial' distribution only by internal
processes. Comparison of the present satellite distribution with that
of Galactic field clusters, with appropriate allowance for recent
destruction, then determines the satellite late accretion rate/fraction
into the galaxy.

Such analyses, now becoming possible, link the evolution of cluster
systems directly with the evolution of a system of individual
clusters, making quantitative the whole life cycle of these stellar
dynamical systems.

\section{Acknowledgement}
It is a pleasure to be part of Ivan King's career
recognition, and to look forwards. I much admire Ivan's achievements,
both scientifically and in his university career, often under
conditions where opportunities were much less open than they are
today. I also much enjoyed working with Ivan on our Saas Fee lecture
course, and the consequent book (Gilmore, King and vanderKruit 1989).

\end{document}